\documentclass[twocolumn]{pasj01}
\Received{}%{yyyy/mm/dd}
\Accepted{}%{yyyy/mm/dd}
\usepackage{multicol}
\usepackage{comment}
\usepackage{url}
\usepackage{lineno} % must be inserted at least since 2021/9/22 to show line numbers

\begin{document}

\title{
%\LETTERLABEL %%% <-- uncomment for LETTER article
%\REVIEWLABEL %%% <-- uncomment for REVIEW article
Subaru/FOCAS IFU revealed the metallicity gradient of a local extremely metal-poor galaxy}

%%% begin:list of authors
% Do NOT capitalize all letters in "textsc".
\author{Yuri \textsc{Kashiwagi}\altaffilmark{1,2,3}%
%\thanks{Present Address is }
}
\email{kasiwagi@km.icrr.u-tokyo.ac.jp}
%\email{yuri0821@ruri.waseda.jp}

\author{Akio K. \textsc{Inoue}\altaffilmark{1,4}}%
\email{akinoue@aoni.waseda.jp}

\author{Yuki \textsc{Isobe}\altaffilmark{2,3}}

\author{Kimihiko \textsc{Nakajima}\altaffilmark{5}}

\author{Masami \textsc{Ouchi}\altaffilmark{5,2,6}}

\author{Shinobu \textsc{Ozaki}\altaffilmark{5}}

\author{Seiji \textsc{Fujimoto}\altaffilmark{7,8}}

\author{Yoshiaki \textsc{Ono}\altaffilmark{2}}

\author{Takashi \textsc{Kojima}\altaffilmark{2,3}}

\altaffiltext{1}{Department of Physics, School of Advanced Science and Engineering, Faculty of Science and Engineering, Waseda University, 3-4-1 Okubo, Shinjuku, Tokyo 169-8555}
\altaffiltext{2}{Institute for Cosmic Ray Research, The University of Tokyo, 5-1-5 Kashiwanoha, Kashiwa, Chiba 277-8582}
\altaffiltext{3}{Department of Physics, Graduate School of Science, The University of Tokyo, 7-3-1 Hongo, Bunkyo, Tokyo 113-0033}
\altaffiltext{4}{Waseda Research Institute of Science and Engineering, Faculty of Science and Engineering, Waseda University, 3-4-1 Okubo, Shinjuku, Tokyo 169-8555}
\altaffiltext{5}{National Astronomical Observatory of Japan, 2-21-1, Osawa, Mitaka, Tokyo 181-8588}
\altaffiltext{6}{Kavli Institute for the Physics and Mathematics of the Universe (WPI), University of Tokyo, Kashiwa, Chiba 277-8583}
\altaffiltext{7}{Cosmic Dawn Center (DAWN), Jagtvej 128, DK2200 Copenhagen N, Denmark}
\altaffiltext{8}{Niels Bohr Institute, University of Copenhagen, Lyngbyvej 2, DK2100 Copenhagen \O, Denmark}

%\author{
%\textsc{},\altaffilmark{2}}
%\altaffiltext{2}{B-Address of Institute}
%\email{bbbbb@xxx.xxx.xx.xx}

%%% end:list of authors

%% `\KeyWords{}' always has to be placed before ``\maketitle''
%%  List of Key Words:  https://academic.oup.com/pasj/pages/Pasj_Keywords

\maketitle

%%%%%%%%%%%%%%%%%%%%%%%%%%%%%%%%%%%%%%%%%%%%%%%%%%%%%%%%%%%%%%%%%%%%%
\begin{abstract}
We present the first measurement of the metallicity gradient in extremely metal-poor galaxies (EMPGs).
With Subaru/Faint Object Camera And Spectrograph (FOCAS) Integral Field Unit (IFU), we have observed a nearby, low-mass EMPG, HSC J1631+4426, whose oxygen abundance and stellar mass are known to be 12+log(O/H) $=6.9$ and $\log_{10}(M_*/{\rm M}_\odot)=5.8$, respectively.
The measured metallicity gradient is $-0.36 \pm 0.04$ dex kpc$^{-1}$ corresponding to $-0.049 \pm 0.006$ dex R$_\mathrm{e}^{-1}$ for the continuum effective radius of $R_\mathrm{e} = 0.14$ kpc. %from Isobe+21 Tab.1 KS2-EMPG, Re = 137^{+0.009}_{-0.007} $
Our observation has successfully demonstrated that three-dimensional spectroscopy with 8m-class telescopes is powerful enough to reveal the metallicity distribution in local EMPGs, providing precious information of the baryon cycle in local analogs of primordial galaxies in the early Universe.
\end{abstract}
%\linenumbers % must be inserted at least since 2021/9/22 to show line numbers
%\clearpage
%%%%%%%%%%%%%%%%%%%%%%%%%%%%%%%%%%%%%%%%%%%%%%%%%%
%%%%%%%%%%%%%%%%%%%%%%%%%%%%%%%%%%%%%%%%%%%%%%%%%%%
%%%%%%%%%%%%%%%%%%%%%%%%%%%%%%%%%%%%%%%%%%%%%%%%%
\section{Introduction}\label{sec1}
Metallicity of galaxies are the key to unveiling their formation and evolution processes (e.g., \cite{pagel1997}).
Metallicity gradient, expressed in dex kpc$^{-1}$ or dex $\rm{R_e}^{-1}$, where ${\rm R_e}$ is the effective radius, is one of the important diagnostics for evaluating gaseous flows in the interstellar medium (ISM) and its mixing processes (e.g., \cite{sharda2021}).
Observations have demonstrated that normal disk galaxies in the local Universe and dwarf galaxies in Local Group have negative metallicity gradient (e.g., \cite{belfiore2017sdss} and references therein). %See bibfile: Leaman+13, Kacharov+17
In the ``inside-out'' galaxy growth scenario in which the star formation begins at the central part of the galaxy and propagates to the outer region, the metallicity gradient is negative and steep in the early stage and flattens as the galaxy grows in size and mass (\cite{wang2019discovery} and references therein). %also in section 1 in X. Ma+17(simulation)
On the other hand, there are also several galaxies that have positive metallicity gradient both in local and distant Universe (e.g., \cite{wang2019discovery, belfiore2017sdss, cresci2010gas}).
The positive gradient observed in high-$z$ galaxies indicates the gas accretion to their central part in the ``cold mode'', with the accretion of pristine gas from the cosmic web (\cite{almeida2013local}).
Since these studies commonly target on galaxies with 12+log(O/H) $\geq 8$, metallicity gradients in metal-poor galaxies are observationally hitherto unknown.

In this Paper, we report a negative metallicity gradient in a local extremely metal-poor galaxy (EMPG), HSC J1631$+$4426, whose metallicity fall on this unexplored range.
EMPGs are defined as galaxies with their oxygen abundance lower than 12+log(O/H) = 7.69, which is equivalent to 10\% solar metallicity (\cite{almeida2015localized}, \cite{kojima2020extremely}, \cite{isobe2020empress}).
They are believed to be the dominant population in the early Universe.
Local EMPGs are of great importance as analogs of high-$z$ galaxies which are difficult to observe directly.
More than 90\% of EMPGs have diffuse structures in their immediate vicinity \citep{isobe2020empress}. %from Isobe+20 section 3.1: 23/27 EMPG have PAGs (> 5sigma i-band magnitude within 10kpc)
Preceding studies suggest that EMPGs are star-forming regions in such diffuse host galaxies whose metallicities are higher than the metal-poor parts, indicating positive metallicity gradients when considered their whole structures (\cite{almeida2013local}, \cite{almeida2016search}, \cite{olmo2017kinematics}).
Our result suggests possible diversity of metallicity gradients in EMPGs.

The rest of this Paper is organized as follows.
In section~\ref{sec2}, observational information and the data reduction process are presented.
In Section~\ref{sec3}, we explain the method of flux measurement and gas-phase oxygen abundance (i.e., metallicity)\footnote{For the sake of simplicity, we refer hereafter to gas-phase oxygen abundance (i.e., not in the stellar atmosphere) as metallicity.)} calculation.
Section~\ref{sec4} describes the measurement of metallicity gradient and discusses its relation with the central metallicity and stellar mass.
Throughout this Paper, a standard $\Lambda$CDM cosmology with parameters of $(\Omega_\mathrm{m},\ \Omega_\Lambda,\ H_0) = (0.286,\  0.714,\ 69.6\ \mathrm{km}\ \mathrm{s}^{-1} \ \mathrm{Mpc}^{-1})$ is adopted.  This cosmology gives a scale of 0.629 kpc per arcsec at $z=0.03125$.  Solar metallicity $Z_{\odot}$ is defined by $12+\mathrm{log(O/H)}=8.69$  \citep{Asplund+09}.

%%%%%%%%%%%%%%%%%%%%%%%%%%%%%%%%%%%%%%%%%%%%%%%%%%%
%%%%%%%%%%%%%%%%%%%%%%%%%%%%%%%%%%%%%%%%%%%%%%%%%%%
%%%%%%%%%%%%%%%%%%%%%%%%%%%%%%%%%%%%%%%%%%%%%%%%%%%
\section{Observation and data reduction}\label{sec2}

Our target, HSC J1631+4426, is identified to be the most metal-poor galaxy with a 1.6\% solar metallicity (i.e., 12+log(O/H) = 6.90;  \cite{kojima2020extremely}).
HSC J1631+4426 is a local dwarf galaxy at redshift $z = 0.03125$, which has been discovered in a metal-poor galaxy survey program named Extremely Metal-Poor Representatives Explored by the Subaru Survey (EMPRESS; \cite{kojima2020extremely}).
Its stellar mass is $\log(M_\star /\rm{M_\odot})=5.8$ and effective radius is measured as $R_\mathrm{e} = 137^{+9}_{-7}$ pc using the Subaru Hyper Suprime-Cam (HSC) \textit{i}-band surface brightness profile \citep{isobe2020empress}.
To reveal the metallicity distribution of the galaxy, we utilize Faint Object Camera And Spectrograph (FOCAS: \cite{kashikawa2002focas}) Integral Field Unit (IFU: \cite{ozakiFOCASIFU}) mounted on Subaru Telescope.
We carried out integral field spectroscopy for the target galaxy, HSC J1631+4426 (RA=16$^{\rm h}$31$^{\rm m}$14$.^{\rm s}$24, DEC=$+$44$^\circ$26$'$04$.''$43), on 2020 March 6 with FOCAS IFU (PI: S. Fujimoto).
We obtained a single 1200-second exposure using the 300B grism and the SY47 filter with the wavelength coverage of 4700--7600 \AA{} and the spectral resolution of  $R\equiv\lambda/\Delta\lambda\sim900$.
An O-type subdwarf, HZ44 (RA=13$^{\rm h}$23$^{\rm m}$35$.^{\rm s}$263, DEC=$+$36$^\circ$07$'$59$.''$55) was also observed as a standard star.
% first submission:HZ44(RA=13$^{\rm h}$23$^{\rm m}$35$.^{\rm s}$37, DEC=$+$36$^\circ$08$'$00$.''$0
During the observation, atmospheric condition was good with the seeing size varying between $0.''6$ and $0.''7$.

To reduce the target and the standard star data, we used the FOCAS IFU pipline software\footnote{\url{https://www2.nao.ac.jp/~shinobuozaki/focasifu/}}.
A detailed explanation of the reduction flow is given in  \citet{ozakiFOCASIFU}.
The size of spatial pixel (called spaxel) and spectral pixel of the final data cube are $0''.435 \times 0''.211$ and 1.34~\AA, respectively.
The total field of view is $10''.0 \times 13''.5$.% should be written in tate x yoko

%%%%%%%%%%%%%%%%%%%%%%%%%%%%%%%%%%%%%%%%%%%%%%%%%%
%%%%%%%%%%%%%%%%%%%%%%%%%%%%%%%%%%%%%%%%%%%%%%%%%
%%%%%%%%%%%%%%%%%%%%%%%%%%%%%%%%%%%%%%%%%%%%%%%%%%%
\section{Analysis}\label{sec3}
We visually identified four emission lines (H$\alpha$, H$\beta$, [O \emissiontype{III}]4959, and [O \emissiontype{III}]5007) in our data cube.
The velocity Full Widths at the Half Maximum (FWHMs) of the lines are $\simeq300$ km s$^{-1}$ (i.e., 5--6 spectral pixels). They are about the velocity resolution: the lines are unresolved. We integrated the data cube over twice of the FWHM for each line to create the velocity-integrated line intensity map.

For each velocity-integration, we subtracted continuum (or sky-residual in spaxels without the object).
The continuum (or sky-residual) level in each spaxel was obtained by integrating the spectra in the wavelength around the emission line over the same number of spectral pixels as that in the line map.
The wavelength range was carefully chosen in order not to include any emission lines.

\begin{figure*}[t]
 \begin{center}
  \includegraphics[width=140mm]{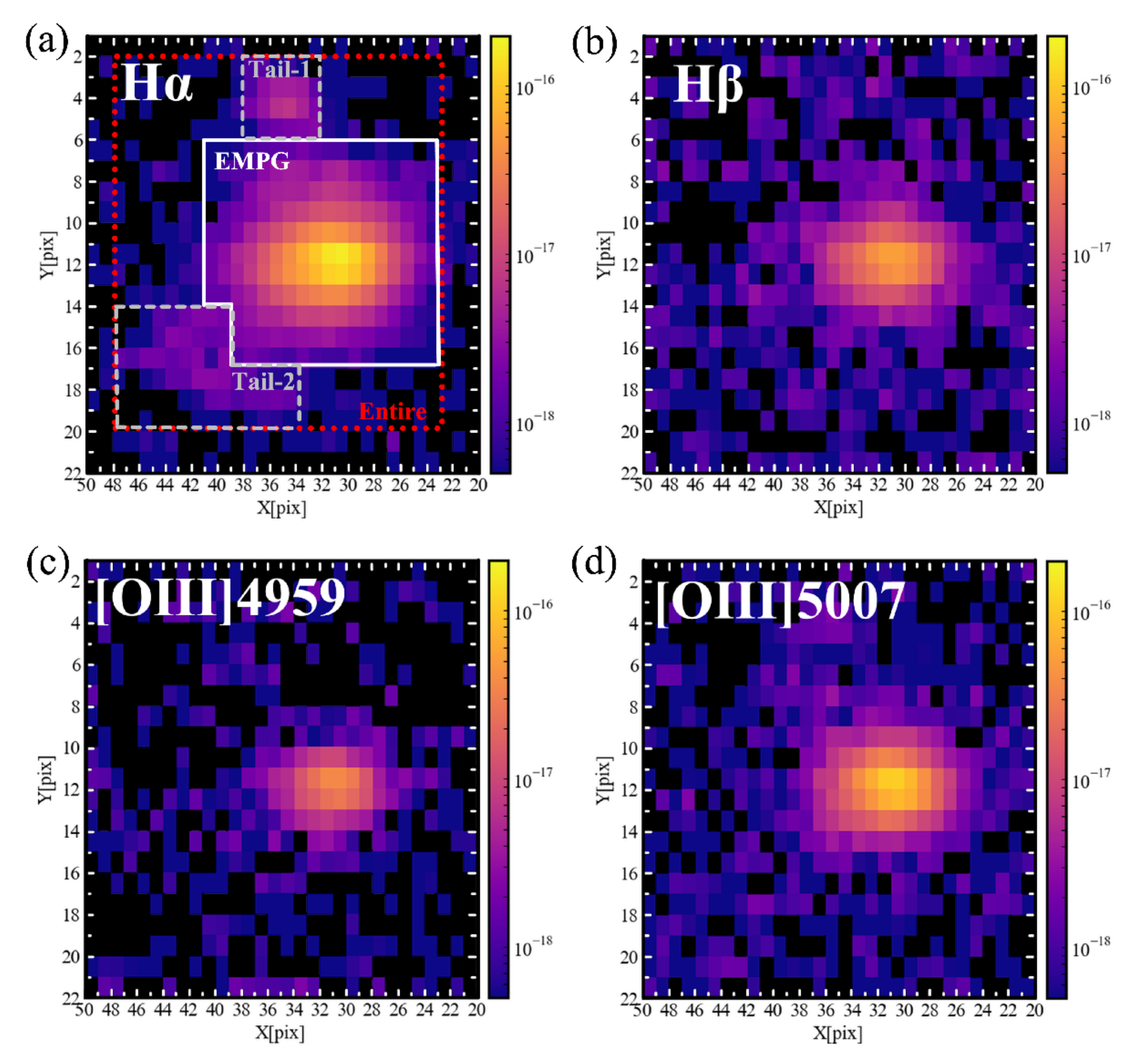}
 \end{center}
\caption{Velocity-integrated line intensity maps in units of $10^{-17}$ erg cm$^{-2}$ s$^{-1}$pix$^{-1}$.  Upward direction and leftward direction of each panel correspond to the north and the east, respectively.  The brightest component of each panel corresponds to the EMPG.  (a)~ The intensity map for H$\alpha$ with the definition of the regions overlaid.
(b-d)~ Same as (a) but for H$\beta$, $\mathrm{[O \emissiontype{III}]4959}$, and $\mathrm{[O \emissiontype{III}]5007}$, respectively.}\label{fig:linemaps}
\end{figure*}

Figure~\ref{fig:linemaps} shows the results.

As shown in Figure~\ref{fig:linemaps} (a) H$\alpha$ map, we defined four regions: the Entire region, the EMPG region, the Tail-1 region, and the Tail-2 region. The latter two regions correspond to the diffuse structures found in the continuum image \citep{isobe2020empress}.
We measured the total line fluxes for each region by summing up the intensity within the region of each line intensity map.
As flux errors, we used RMS per spaxel calculated from the spaxels outside of the Entire region and scaled it by the square-root of the number of the spatial pixels.
We also measured the upper limits of the line fluxes of [N\emissiontype{II}]6584, [S\emissiontype{II}]6717, and [S\emissiontype{II}]6731.
We corrected the total fluxes for the dust extinction within the Milky Way by utilizing Galactic Dust Reddening and Extinction Service\footnote{\url{https://irsa.ipac.caltech.edu/applications/DUST/}} and the \citet{cardelli1989relationship} extinction curve.
Visual extinction by the Milky Way toward the target galaxy is $A_{\mathrm{MW}}(V)=0.0237$ mag.
The corrected line fluxes and upper limits are listed in Table~\ref{tab: measurement}.

For the dust attenuation in the target galaxy, we estimated the color excess $E(B-V)$ for each region based on the H$\alpha$/H$\beta$ line ratio, the Balmer Decrement, under the assumptions of the \citet{cardelli1989relationship} extinction curve, the electron temperature $T_\mathrm{e} = 25000$ K measured by [O\emissiontype{IIII}]4363 \citep{kojima2020extremely}, and the case B approximation (\cite{osterbrock1989}).
The obtained $E(B-V)$ values are listed in Table~\ref{tab: measurement}.
Since we obtained non-zero $E(B-V)$ only for the Tail-1 region, we applied the internal dust attenuation correction only for that region, assuming the \citet{cardelli1989relationship} extinction curve.

The gas-phase metallicity, 12+log(O/H) for each region was estimated from the $R_3$ index: $R_3 (\equiv \mathrm{[O~\sc{III}]5007} / \rm{H\beta})$ \citep{maiolino2019re}.
Since the relation between $R_3$ and 12+log(O/H) is known to be a binary function, there can be high and low 12+log(O/H) branches for a value of $R_3$.
To know the more likely case, we also examined the $N_2$ index: $N_2 (\equiv \mathrm{[N \emissiontype{II}]6584} / \mathrm{H\alpha})$.
Although the [N \emissiontype{II}]6584 line is not detected in the galaxy, its upper limit can be useful to reject the high 12+log(O/H) solution.

First, we calculated the $R_3$ and $N_2$ indices for each region, using the dust-corrected total line fluxes and upper limits.
Next, we used the empirical relation by \citet{curti2020mass} calibrated for the $N_2$ index to obtain upper limits of 12+log(O/H).
%from the observed $N_2$ upper limits.
We found that all regions should have 12+log(O/H) $ < 7.728$--$8.550$
which is consistent with the EMPG nature \citep{kojima2020extremely}. %the min and max of the upper limit 12+log(O/H)
\citet{curti2020mass} also present the $R_3$ calibration, which is limited to 12+log(O/H) $> 7.6$, while the target galaxy is an EMPG  \citep{kojima2020extremely}.
We then used the theoretical models by \citet{inoue2011test} which cover a very wide range of 12+log(O/H).

The models were calculated by \textsc{CLOUDY} \citep{ferland1998} for the cases with gas-phase metallicity ($\mathrm{log}_{10}(Z/Z_\odot)=-\infty, -5.3, -3.3, -1.7, -0.7, -0.4,$ and $0.0$), ionization parameter ($-3\leq \log_{10}U \leq -1$), and hydrogen number density ($0\leq \log_{10}(n_{\rm H}/{\rm cm^{-3}}) \leq 2$).
They also changed the input stellar spectra depending on the metallicity (gas and stellar metallcities were assumed to be the same).
Finally, they presented the emission line intensities normalized by H$\beta$ as a function of metallicity, averaging over the 25 different sets of ($U$, $n_{\rm H}$) in each metallicity case.
They noted that the standard deviations of the line ratios relative to H$\beta$ are 5--25\%.
The obtained average line ratios are consistent with observations as shown in their Figure~1 for some strong emission lines including [O~\emissiontype{III}] lines.
Since the metallicity grid of the line ratios was sparse,
we used the following interpolation function:
\begin{eqnarray}\label{eq:interpolation}
    \log_{10} (R_3) =
    \left\{
    \begin{array}{llrr}
    9.4272\times10^{-1}x -6.5028 \hspace{8pt}(\mathrm{for} \hspace{1pt} x\leq 6.9910 ) \\
    -5.2795\times10^{-4}x^5 +3.7631\times10^{-3}x^4 \\
    \hspace{10pt} +3.2170\times10^{-2}x^3 -2.0897\times10^{-2}x^2 \\
    \hspace{20pt} -1.8000x +3.2803 \hspace{8pt}(\mathrm{for}\hspace{1pt} x\geq 6.9910 )
    \end{array}\right.
\end{eqnarray}
where $x$ = 12+log(O/H).
%\textcolor{red}{
%The equation \ref{eq:internpolation} is derived by interpolating the data points of \citet{inoue2011test} using the method Polyfit in Python module numpy.
%}
The obtained 12+log(O/H) for each region are summarized in Table~\ref{tab: measurement}.

%WCS axis

\begin{table*}[h]
    \tbl{Measurement of emission lines, color excess, and metallicity for the four regions\footnotemark[$*$] }{
    \scalebox{0.9}{
    \begin{tabular}{cccccccccc}
    \hline
     & H$\beta$ & [O \emissiontype{III}]4959 & [O \emissiontype{III}]5007 & H$\alpha$ &[N\emissiontype{II}]6584 & [S\emissiontype{II}]6717 & [S\emissiontype{II}]6731 & $E(B-V)$ & 12+log(O/H)\\
    \hline
    Entire	&	953.5	$\pm$	16.7	&	396.6	$\pm$	15.1	&	1467.1	$\pm$	13.9	&	2493.7	$\pm$	11.1	&$<$	31.6	&$<$	189.0	&$<$	1484.9	&$<$	0.0503 &$7.101	^{+0.010}_{-0.010}$ \\
    EMPG	&	867.6	$\pm$	11.2	&	427.5	$\pm$	10.0	&	1408.6	$\pm$	9.3	&	2367.6	$\pm$	7.4	&$<$	21.1	&$<$	126.0	&$<$	990.0	&$<$	0.0338 &$7.128	^{+0.007}_{-0.007}$\\
    Tail-1	&	15.0	$\pm$	3.9	&$<$ 10.4   &	21.4	$\pm$	3.3	&	57.7	$\pm$	2.6	&$<$	7.4	&$<$	44.6	&$<$	350.0	&0.349	$\pm$	0.114 &$7.043	^{+0.133}_{-0.168}$\\
    Tail-2	&	27.0	$\pm$	6.4	&$<$	17.2			&	20.8	$\pm$	5.3	&	65.7	$\pm$	4.3	&$<$	12.0	&$<$	72.1	&$<$	567.1	&$<$	0.789 &$6.778	^{+0.138}_{-0.197}$\\
    \hline
\end{tabular}}}\label{tab: measurement}
\begin{tabnote}
      \footnotemark[$*$]  Measurement for the total fluxes after the correction for the Milky Way dust attenuation, $E(B-V)$ in the target galaxy, and 12+log(O/H). The observed total flux within each region in units of 10$^{-18}$ erg cm$^{-2}$ s$^{-1}$.  Upper limits are given at the 3$\sigma$ level.
\end{tabnote}
\end{table*}

\begin{figure}[t]
 \begin{center}
  \includegraphics[width=70mm]{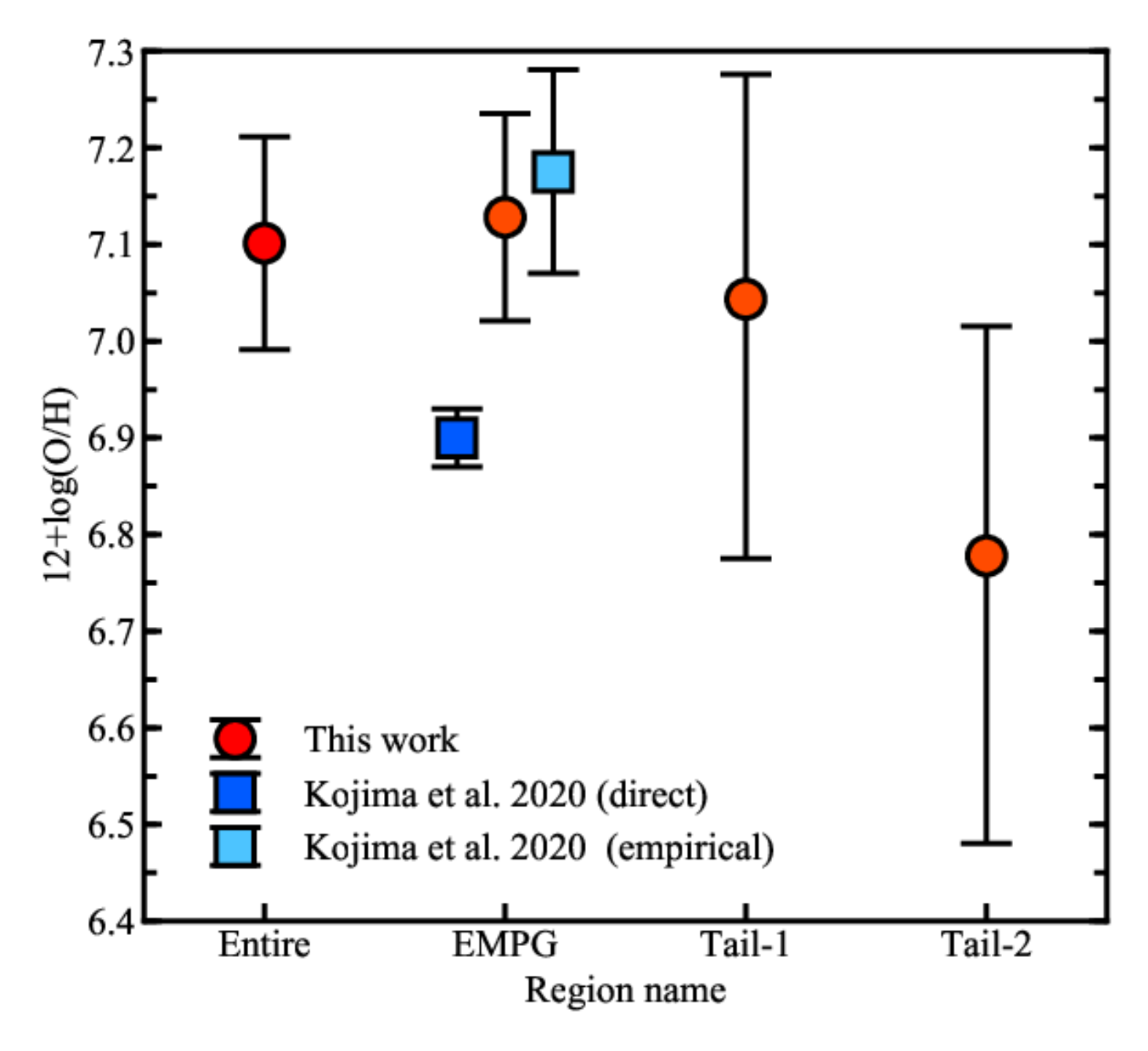}
 \end{center}
\caption{Comparison of 12+log(O/H). The red circles represent our oxygen abundances. The blue and cyan squares represent those reported by Kojima et al.~(2020a) based on a {\it direct} method and an empirical method, respectively.  We note that systematic uncertainties of $\pm$0.1 are added to the data, except for the direct method one.}\label{fig:metallicity plot}
\end{figure}

Figure \ref{fig:metallicity plot} shows our 12+log(O/H) for each region with error-bars, plotted along with those reported by  \citet{kojima2020extremely}.
We added systematic uncertainty of $\pm0.1$ to the results based on empirical calibrations because such dispersion exists in these calibrations (e.g., \cite{maiolino2019re}).
The oxygen abundances for the Entire, EMPG, and Tail-1 regions are consistent with each other within the uncertainty when the systematic errors are considered.
The oxygen abundance of the Tail-2 region may be lower than the other three regions although the uncertainty is large.

There is a systematic offset of our 12+log(O/H) values from that obtained by a direct temperature method in  \citet{kojima2020extremely}.
We could not detect the [O\emissiontype{III}]4363 in our data cube because of the short exposure.
It is a future work to apply the direct temperature method to the target galaxy in a spatially resolved way after taking a deeper data set (see also a discussion below).

%%%%%%%%%%%%%%%%%%%%%%%%%%%%%%%%%%%%%%%%%%%%%%%%%%
\section{Metallicity gradient measurement}\label{sec4}
To derive metallicity gradient, we created metallicity map by calculating 12+log(O/H) for each spaxel based on the $R_3$ index as in section~\ref{sec3}.
Since the H$\beta$ line flux is required in the denominator of the $R_3$ index, we restricted ourselves to the spaxels where the S/N of H$\beta$ is larger than 3, which are confined only in the EMPG region.
The line fluxes corrected only for the Milky Way dust extinction were used for the calculation because $E(B-V)$ in the EMPG region is consistent with zero (Table~\ref{tab: measurement}).

In this Paper, we present the metallicity gradient purely observationally and do not consider any geometric model to correct for the inclination effect.
Since our target galaxy is not edge-on, the projection effect on the distance may not be too large.
We assumed the center of the galaxy to be the intensity peak position of the H$\alpha$ map and calculated the projected distance of each spaxel from the center.
Note that the kinematic center of the H$\alpha$ emission is consistent with its intensity peak (Isobe et al. in prep).

\begin{figure}[t]
 \begin{center}
  \includegraphics[width=70mm]{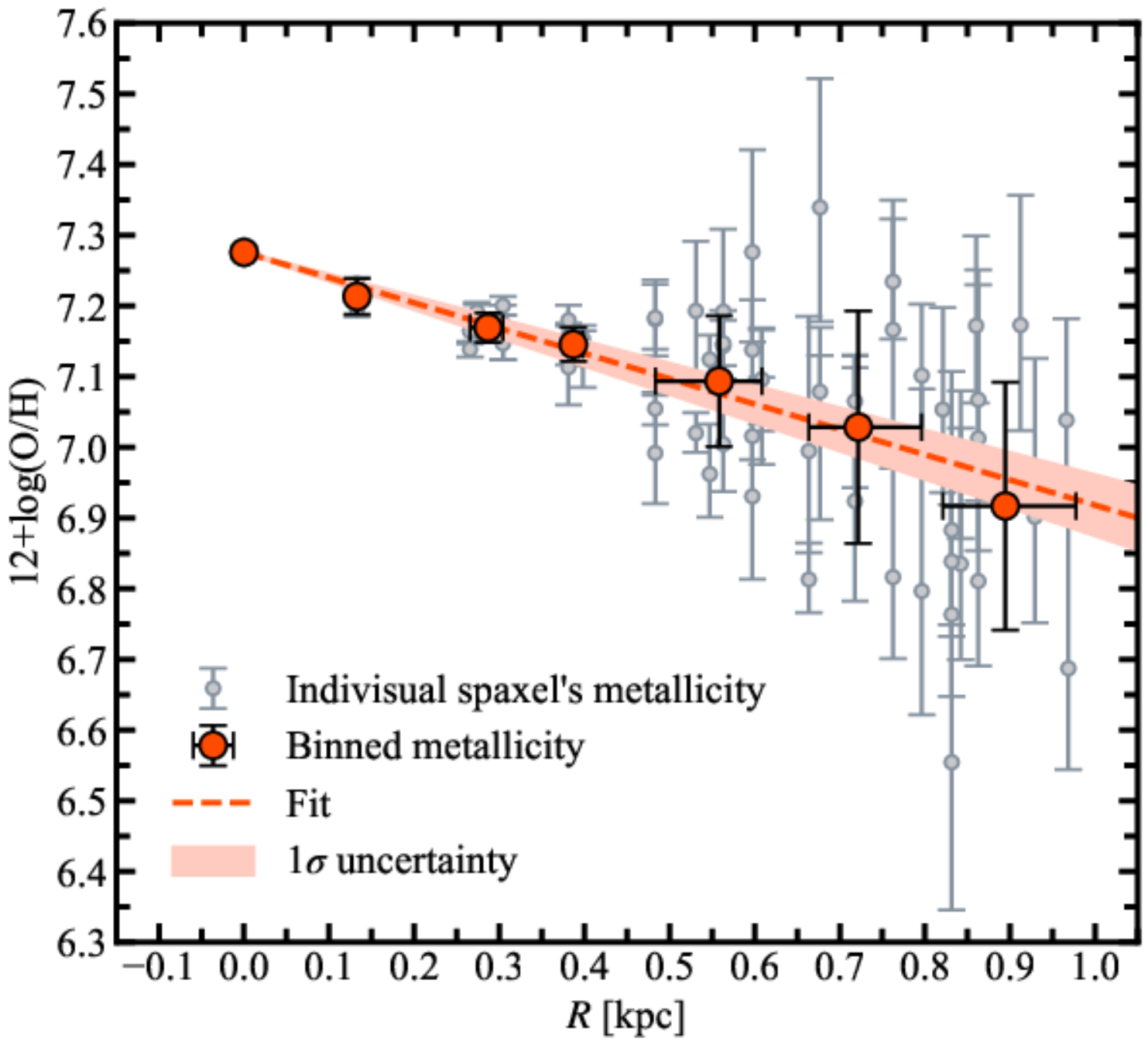}
 \end{center}
\caption{Radial profile of metallicity in the EMPG region of J1631+4426.  The grey circles represent the metallicities of individual spaxels with the observational errors.  The red circles represent the binned metallicities with the errors indicating the sample standard deviations along the both axes.  The red dashed line is the linear fit of the binned metallicities by a $\chi^2$ method.}
\label{fig: metallicity gradient}
\end{figure}

Figure~\ref{fig: metallicity gradient} shows the obtained radial profile of the metallicity in the EMPG region.

It is important to consider the beam smearing effect on metallicity gradient measurements \citep{yuan2013}.
Since our metallicity measurements reach a scale twice larger than the seeing size ($\approx 0.''7=0.44$ kpc), the beam smearing effect would not be large.

We fitted the binned data and their standard deviations with a linear function by a $\chi^2$ method,
varying both the central metallicity and the gradient.
We obtained $-0.36 \pm 0.04$ dex kpc$^{-1}$ as the metallicity gradient as well as the central metallcity of 12+log(O/H) $=7.276\pm0.001$.
Note that the observational seeing of $0.''7$ does not affect the linear function fit because a linear function is conserved by a Gaussian convolution. We also obtained a consistent gradient value even when we divided the data into two groups (i.e., the inside and the outside of $0.44$ kpc).

%FIG4
\begin{figure*}[t]
 \begin{center}
  \includegraphics[width=80mm]{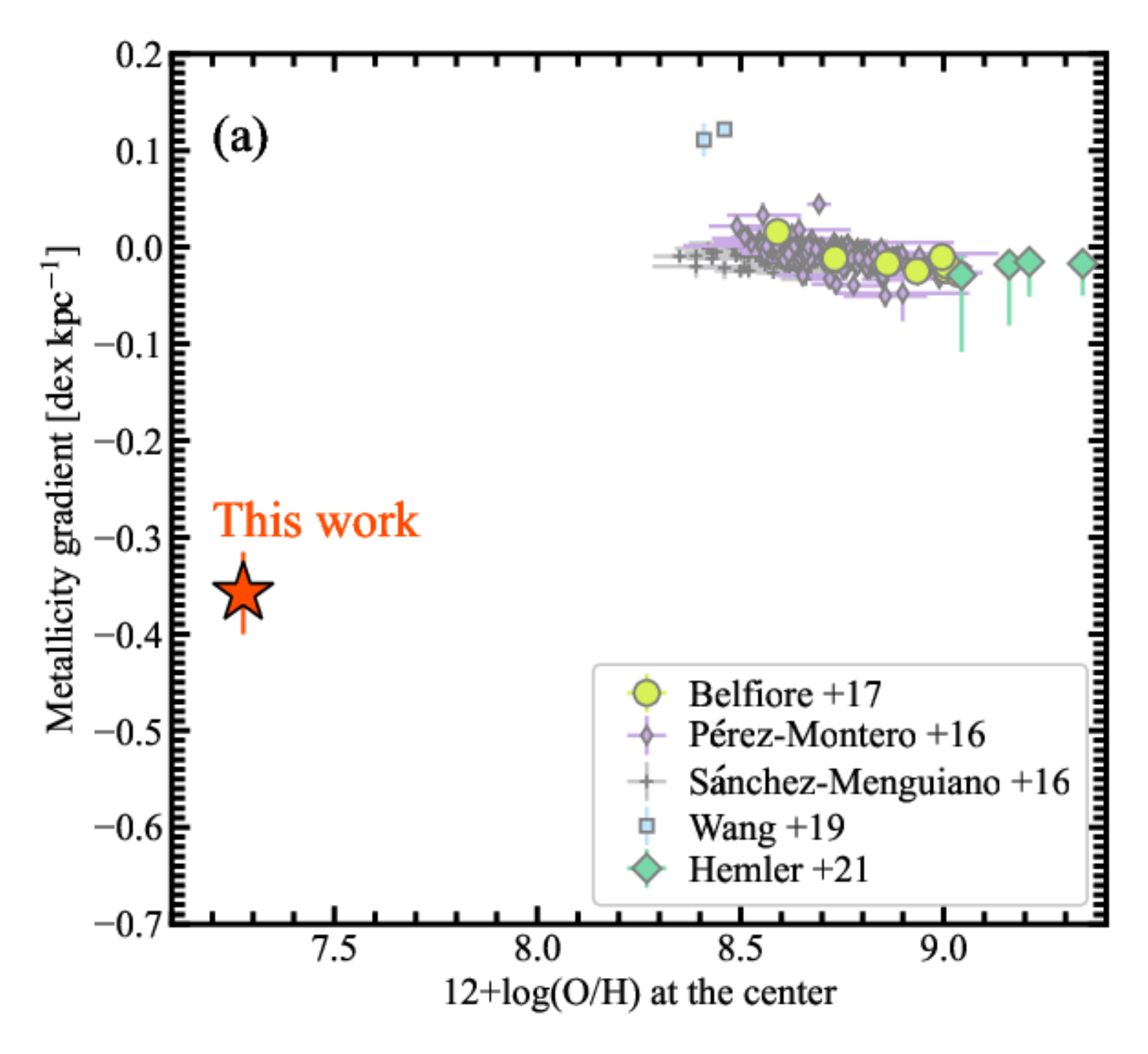}
  \includegraphics[width=80mm]{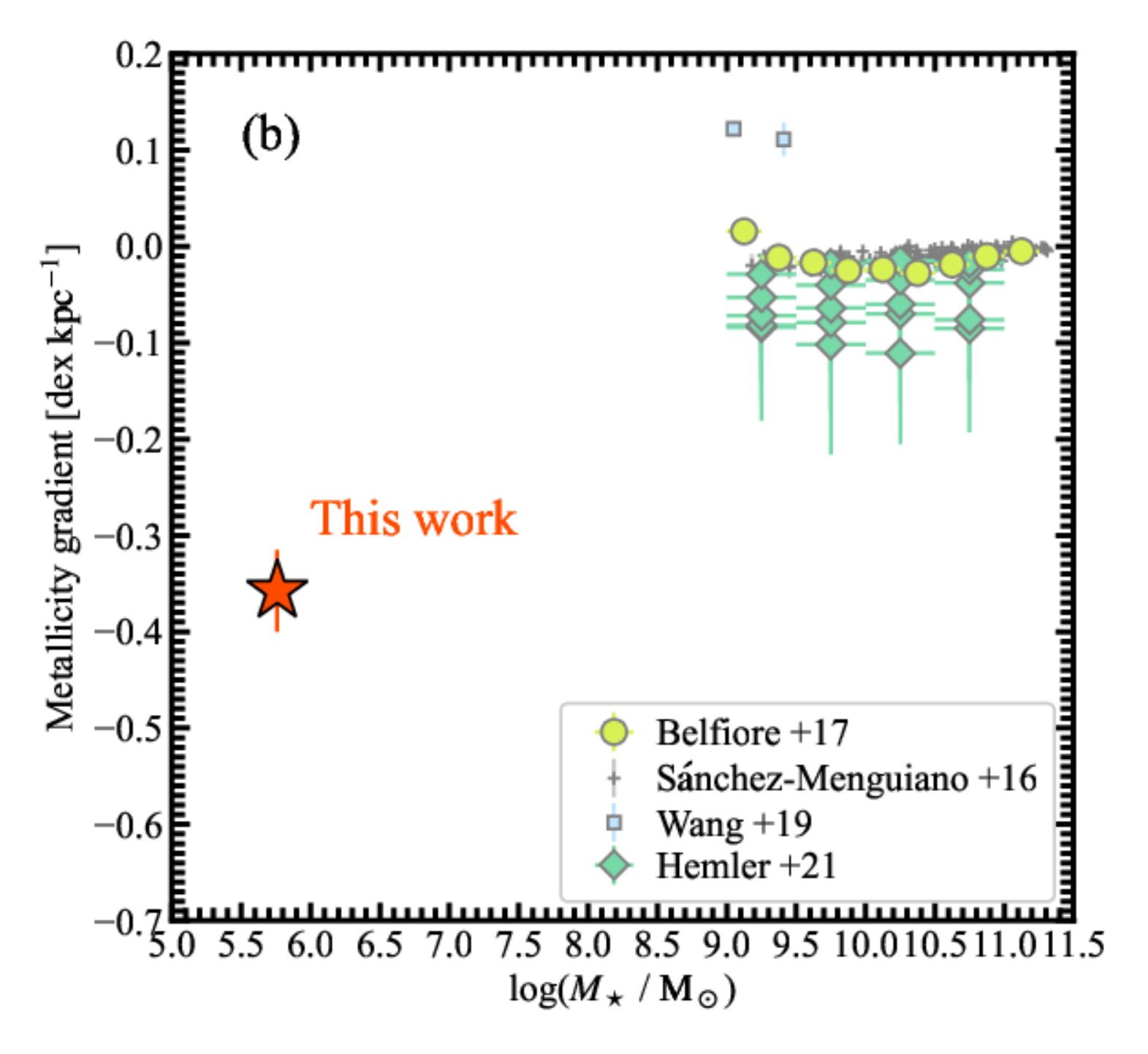}
 \end{center}
\caption{
(a) The dependence of metallicity gradient on central metallicity. The red star represents this work.
The light green circles indicate the median metallicities in 0.25 dex stellar mass bin of SDSS galaxies with $\log(M_\star / {\rm M}_\odot) = 9.0 - 11.5$ at z $\sim$ 0 \citep{belfiore2017sdss}.
The purple, skewed diamonds indicate the median metallicities representing $\log(M_\star / {\rm M}_\odot) = 9.0, 9.5, 10.0, 10.5$ CALIFA galaxies at z $\sim$ 0 \citep{perez2016dependence}.
The grey plus marks represent 122 CALIFA galaxies with $\log(M_\star / {\rm M}_\odot) = 9.18 - 11.31$ at z $\sim$ 0 \citep{sanchez2016shape}.
The blue squares show the positive metallicity gradients of two dwarf galaxies at $z \sim 2$ \citep{wang2019discovery}.

The green diamonds are the TNG-50 simulated star-forming galaxies with $\log(M_\star / {\rm M}_\odot) = 9 - 11$ at $z = 0$ by \citet{hemler2021gas}.

(b) Same as (a) but for the dependence on the stellar mass.

The redshift of the simulated galaxies by \citet{hemler2021gas} (green diamonds) for each mass bin $\log(M_\star / \rm{M_\odot})=9.25, 9.75, 10.25, 10.75$ is $z = 0, 0.5, 1, 2, 3$ from top to bottom, respectively.

}
\label{fig:gradient vs.}
\end{figure*}

%%%%%%%%%%%%%%%%%% DISCUSSION %%%%%%%%%%%%%%%%%
To evaluate the properties of our galaxy, we compare our result with previous studies.
Figure \ref{fig:gradient vs.} shows the metallicity gradient as functions of the central metallicity (a) and the stellar mass (b), respectively.
Our work successfully reports the metallicity gradient in a considerably metal-poor, low-mass galaxy with $\log(M_\star / \rm{M_\odot}) \sim 6 $ for the first time.

The metallicity gradient of our EMPG in units of dex $\mathrm{kpc}^{-1}$ is considerably steep compared to those almost flat observed in local high-mass galaxies ($\log(M_\star / \rm{M_\odot}) \sim 9$).
This steep gradient can be interpreted by a simple chemical evolution model as follows.

A simple closed-box model approximately gives the time evolution of oxygen abundance as $Z_{\rm O}\approx y_{\rm O} \left(\frac{t}{\tau_{\rm SF}}\right)$ when $t<\tau_{\rm SF}$ \citep{pagel1997}, where $y_{\rm O}$ is the stellar yield of oxygen and $\tau_{\rm SF}$ is the time-scale of star formation.
This equation can be reduced to
$\left(\frac{\Delta t_{\rm{chem}}}{\tau_{\rm SF}}\right) = \rm{ln10} \left(\frac{Z_{\rm O}}{y_{\rm O}}\right)\left|\frac{\Delta 12+{\rm log}({\rm O/H})}{\Delta R}\right| \Delta R \approx 0.02 \left(\frac{\Delta R}{\rm 1~kpc}\right)$
when we adopt $Z_{\rm O}\approx0.02 Z_{\rm O,\odot}$  \citep{kojima2020extremely}, $Z_{\rm{O,\odot}}=5.7\times10^{-3}$  \citep{Asplund+09}, $y_{\rm{O}}=0.004$  \citep{meynet2002,almeida2015localized}, and our metallicity gradient of $\left|\frac{\Delta12+{\rm{log}}({\rm O/H})}{\Delta R}\right| = 0.36$ kpc$^{-1}$.
Therefore, the observed metallicity gradient can be realized quickly compared to the galaxy evolution (i.e., $\tau_{\rm{SF}}$).
On the other hand, the ISM mixing time-scale is $\Delta t_{\rm{mix}}=\left(\frac{\Delta R}{v_{\rm{mix}}}\right)$ with the mixing velocity $v_{\rm{mix}}$.
The presented data cube of the H$\alpha$ line indicates a velocity gradient of $\simeq30$ km s$^{-1}$.
Assuming $v_{\rm{mix}}$ to be the same order of that velocity, we find $\left(\frac{\Delta t_{\rm chem}}{\Delta t_{\rm mix}}\right)=0.7\left(\frac{\tau_{\rm SF}}{\rm 1~Gyr}\right)\left(\frac{v_{\rm mix}}{\rm 30~km~s^{-1}}\right)$.
If $\tau_{\rm{SF}}\lesssim1$ Gyr, a metallicity gradient produced by the chemical enrichment can be kept against the ISM mixing.
The observed steep negative gradient indicates inside-out star formation in the EMPG and its inefficient ISM mixing due to a slow turbulent velocity induced by a shallow gravitational potential.

On the origin of the metal-poor gas of EMPGs, \citet{almeida2015localized} suggest that the cold, metal-poor gas infall from the cosmic web dilutes the metallicity of the central part and triggers star formation there.
They also report that diffuse structures have $\sim1$ dex higher metallicity than the star-forming region of EMPGs. %(section \ref{sec1}).
Therefore, the metallicity gradient of their EMPGs seem positive when we consider the whole system including the diffuse structures.
Positive metallicity gradients are also observed in some local and high-$z$ galaxies \citep{cresci2010gas,sanchez2016shape,wang2019discovery}.
On the other hand, metallicites of the two Tail regions in our galaxy are similar to or lower than the mean of the EMPG region (figure~\ref{fig:metallicity plot}) and are likely to be lower than the centeral metallicity of the EMPG region.
Although the distances of these regions from the EMPG center are about $\sim 2$ kpc, %15 $R_{\rm{e}}$
out of the range of the current analysis shown in Figure~\ref{fig: metallicity gradient}, the galaxy we discuss here probably do not have any positive gradient even if we include the Tail regions.

A caveat of the analysis in this Paper is possible radial dependence of the nebular parameters.
We have implicitly assumed radial constancy of nebular parameters by using the empirical calibration formula.
For example, if the ionization parameter changes radially, the empirical $R_3$-index method we used may suffer from larger uncertainties caused by missing information of the O$^+$ amount traced by the [O\emissiontype{II}]3727 lines.
A potential more serious case is that the electron temperature is as high as $T_e=25000$~K, obtained from [O\emissiontype{III}]4363  \citep{kojima2020extremely}, only in the central part of the EMPG and decreases along the radial distance.
This case may lead to a $\sim0.2$ dex lower metallicity (figure~\ref{fig:metallicity plot}), only in the central part because the same $R_3$ index gives a lower 12+log(O/H) value for higher temperature.
As a result, the gradient would become shallower than that derived here.
Examining these points require a shorter wavelength coverage for [O\emissiontype{II}]3727 or much deeper data cube for faint [O\emissiontype{III}]4363, which is a future work.

Another caveat is the contribution of diffuse ionized gas (DIG) to the emission lines \citep{zhang2017dig,sanders2017,sanders2021}.
The half-light radius of H$\alpha$ emission is measured at $R_{\rm e,H\alpha}=0.''61=380$ pc from the line map, which is about three times larger than $R_{\rm e}=140$ pc of the stellar component measured in the HSC $i$-band image \citep{isobe2020empress}.
In the $GALEX/NUV$ image, the galaxy is detected but not resolved, providing no spatial information of the young stellar component.
If we assume the $i$-band size to be the size of the young stellar component, it is significantly smaller than the spatial extension of the H$\alpha$ emission.
A part of H$\alpha$ emission may come from DIG.
Indeed, the DIG fraction is estimated at $\sim 0.5$ based on equation~(24) in \citet{sanders2017} from the mean H$\alpha$ surface brightness of $\Sigma_{\rm H\alpha}=L_{\rm H\alpha}/2\pi R_{\rm e,H\alpha}^2=6\times10^{39}$ erg s$^{-1}$ kpc$^{-2}$ in our line map.
On the other hand, the metallicity estimated by the $R_3$ index, which we adopted here, may not be affected by the DIG contribution because the $R_3$ indices of HII regions and DIG are similar \citep{zhang2017dig,sanders2017}.
In addition, the DIG correction for the strong line methods for metallicity tends to be small at lower metallicity \citep{sanders2021}, suggesting the DIG effect on our measurements to be small.

%%%%%%%%%%%%%%%%%%%%%%%%%%%%%%%%%%%%%%%%%%%%%%%%%%%
\begin{ack}
We would like to thank anonymous reviewer for insightful comments that are helpful for us to improve the quality and clarity of the Paper.
This work was supported by the joint research program of the Institute for Cosmic Ray Research (ICRR), University of Tokyo.
The Cosmic Dawn Center is funded by the Danish National Research Foundation under grant No.~140.
Takashi Kojima was supported by JSPS KAKENHI Grant Number 18J12840.
We would like to thank the staffs of the Subaru Telescope for their help with the observation.
This research has made use of the NASA/IPAC Infrared Science Archive, which is funded by the National Aeronautics and Space Administration and operated by the California Institute of Technology.
\end{ack}

%%%%%%%%%%%%%%%%%%%%%%%%%%%%%%%%%%%%%%%%%%%%

\end{document}